**Risperidone response in patients with schizophrenia drives DNA methylation changes in the immune and neuronal systems**

Ana Lokmer, Ph.D.[q,b]*[1], Charanraj Goud Alladi, Ph.D.[c1], Réjane Troudet, Pharm.D., Ph.D.[a,b], Delphine Bacq-Daian.[d], Anne Boland-Auge, Ph.D.[d], Violaine Latapie[a,b], Jean-François Deleuze, Ph.D.[d], Ravi Philip RajKumar, M.D.[e], Deepak Gopal Shewade, M.Pharm., Ph.D.[f], Frank Bélivier, M.D., Ph.D.[b,c,g], Cynthia Marie-Claire, Ph.D.[c], and Stéphane Jamain, Ph.D.[a,b,]

a Univ Paris Est Créteil, INSERM, IMRB, Translational Neuropsychiatry, Créteil, France

b Fondation FondaMental, Créteil, France

c Université de Paris, INSERM UMR-S 1144, Optimisation Thérapeutique en Neuropsychopharmacologie (OTeN), Paris, France

d Université Paris-Saclay, CEA, Centre National de Recherche en Génomique Humaine (CNRGH), 91057, Evry, France

e Department of Pharmacology, Jawaharlal Institute of Postgraduate Medical Education and Research, Puducherry-605006, India

f Department of Psychiatry, Jawaharlal Institute of Postgraduate Medical Education and Research, Puducherry-605006, India

f Centre National de Recherche en Génomique Humaine (CNRGH), Institut de Biologie François Jacob, CEA, Université Paris-Saclay, Evry, France

g Hôpitaux Lariboisière-Fernand Widal, GHU APHP Nord, Département de Psychiatrie et de Médecine Addictologique, Paris, France

1 Equal contributions.

*Corresponding authors:

Dr Ana Lokmer

Univ Paris Est Créteil, Inserm U955, IMRB, Neuropsychiatrie Translationnelle

Faculté de Médecine, 8 rue du Général Sarrail, 94010 Créteil cedex, France.

Phone: +33 1 49 81 37 78

E-mail: ana.lokmer@mnhn.fr

**Abstract**

**Background:** The choice of efficient antipsychotic therapy for schizophrenia relies on a time-consuming trial-and-error approach, whereas social and economic burden of the disease calls for faster alternatives. **Material & methods:** In search for predictive biomarkers of antipsychotic response, we analyzed blood methylomes of 28 patients before and four weeks into risperidone therapy. **Results:** We identified several CpGs exhibiting response-specific temporal dynamics in otherwise temporally stable methylomes and observed noticeable global response-related differences between good and bad responders. These were associated with the genes involved in immunity, neurotransmission and neuronal development. Polymorphisms in many of these genes were previously linked with schizophrenia etiology and antipsychotic response. **Conclusion:** Antipsychotic response seems to be shaped by both stable and medication-induced methylation differences.

**Plain Language Summary**

The most common way to treat schizophrenia is antipsychotic medication. However, not all antipsychotics work for all patients. The only way to find a suitable antipsychotic is to prescribe one and wait, sometimes for months, to see if it works. Finding an alternative to this "trial & error" method would help reduce patient suffering and costs for healthcare systems. The idea is to look in the DNA of our blood cells for the specific marks that can change in response to our lifestyle or health condition. These marks could help us predict how the patients will react to the drug. In other words, they can serve as biomarkers of antipsychotic response. So we examined the blood of schizophrenia patients before and four weeks after starting the medication. We found that the patients who did not respond well to the drug had different marks on the genes involved in immune defense and nervous system functioning. Some of these genes also play role in the development of schizophrenia, whereas others can directly affect what happens to the drug in the patient's body. Still, we could not predict how the patients will react with certainty, because we examined only 28 patients. However, we found valuable targets for future research.

**Tweetable Abstract**

Blood methylomes of schizophrenia patients suggest the role of immune and neuronal systems in antipsychotic treatment response and reveal potential targets for antipsychotic response biomarker research.

**Summary Points**

- Trial & error method of choosing an efficient antipsychotic treatment for schizophrenia patients prolongs suffering and places considerable burden on healthcare systems.
- Blood represents an easily accessible source of potential biomarkers predictive of antipsychotic response.
- We examined blood methylomes of good and bad responders to risperidone treatment, just before and four weeks after starting the therapy to look for predictive biomarkers of antipsychotic response.
- Blood methylomes differ slightly between the good and bad responders and are overall temporally stable, but we did observe some response-specific shifts following the treatment.
- Differentially methylated genes were mostly linked to the genes involved in immunity, neurotransmission and neuronal functioning and development.

- Some of the differentially methylated genes, such as *CACNA2D2* or *NAV*, have been previously implicated in schizophrenia risk.
- We also observed differential methylation of glutathione S-transferase *GSTM1* and *MGST1* genes, whose polymorphisms were shown to affect antipsychotic treatment response and their side effects.
- Although we could not identify definitive antipsychotic response biomarkers in our small cohort, our results suggest potential targets for the future response biomarker research.

 Introduction

Schizophrenia (SCZ) is a heritable psychiatric disorder affecting over 20 million individuals worldwide [1] and ranking among the top 20 causes of disability according to the Global Burden of Disease Studies [2]. Its big impact on health, social and economic systems can be explained by early onset of its severe symptoms, its chronic nature and co-occurrence with various somatic diseases, leading to a life-long productivity loss, mental and physical suffering, and reduced life expectancy [1,3]. Development of efficient treatments for SCZ is therefore of paramount importance for the affected individuals and the society as a whole.

Despite the decades of intense research and development of new drugs and treatments following the introduction of the first antipsychotic (AP) medication more than 60 years ago, only a minority of SCZ patients (median of 13.5%) attain durable recovery, i.e. clinical remission and good social functioning for a minimum of two years [4]. The use of antipsychotics is not only associated with severe adverse effects, but also plagued with a varying and unpredictable efficacy in alleviating SCZ symptoms [5,6]. Due to the low predictability of AP treatment success, trial-and-error remains the standard method for therapy adjustment, often taking weeks or months to select a suitable AP and thus prolonging patients' suffering. The need for early and accurate prediction of patient response to medication is further accentuated by the fact that early treatment start is associated with better clinical outcomes [7]. Although the traits such as age at disease onset, ethnicity and baseline symptom severity may help predict the AP response to a certain extent (review in [8]), we recently showed that this prediction was improved by combining clinical data with biological markers [9,10]. Indeed, a considerable individual phenotypic and matching genotypic variability observed in SCZ patients suggests that personalized medicine, including the identification of predictive biomarkers, holds the key to successful SCZ treatment [11].

Most of the research regarding the response to antipsychotics has so far focused on the underlying genetic and genomic variation (reviews in [12,13]). Pharmacogenetic and pharmacogenomic studies identified a number of candidate polymorphisms in the genes involved in dopaminergic and serotonergic system, glutamate signaling [14] as well as various other genes that may serve as predictive biomarkers of antipsychotic response [12]. Interestingly, Ruderfer et al. [15] found an overlap between the SNPs associated with SCZ risk and AP efficacy, not only suggesting the existence of shared mechanisms between SCZ pathogenesis and treatment response, but also demonstrating that the SCZ treatment response is polygenic and complex in nature just as the disease itself. As the treatment outcome is influenced not only by genetics, but also by its interplay with environmental factors, gene expression (GE) profiles may provide more specific information about the AP efficacy than genetic data. Comparisons of GE profiles between healthy controls and patients before and after treatment suggest reversal to "normal" profiles in the treated patients [16,17]. However, there is virtually no overlap between the genes identified in different studies. The reasons for this may include various factors, such as small sample size, omission of important covariates, large individual variation of GE profiles as well as analysis-related technical issues (review in [18]). Similarly, the only two studies [10,19] explicitly trying to predict the treatment outcome from the baseline GE profiles identified different sets of genes as candidate predictive biomarkers, highlighting once more the complex nature of AP response and the role of unknown individual and environmental factors.

Similar to GE, epigenetic marks reflect both genetic and environmental influences, while being more temporally stable and thus likely representing a more reliable source of predictive biomarkers [20]. Although there is a considerable amount of research regarding the epigenetic mechanisms involved in SCZ (review in [21]), most of the studies so far have not been designed to differentiate between the epigenetic signatures of SCZ pathology and medication use, with very few specifically addressing the relationship between DNA methylation (DNAm) and AP treatment response heterogeneity (reviews in [22–24]). A single study up to date using a repeated measures design and a genome-wide approach to search for CpG islands predictive of AP response remains that of Rukova et al. [25]. Rukova et al. identified nine differentially methylated regions (DMRs) that distinguished good from bad responders at the baseline, but only in men. This example highlights the need for detailed patient description in the quest for AP response predictive biomarkers and this holds especially true for traits with a known influence on SCZ risk, antipsychotic response and DNAm profiles, such as age, sex or tobacco use [26–31].

In order to improve our understanding of the link between DNAm profiles and variation in antipsychotic treatment response, we analyzed 56 blood methylomes from 28 patients before and four weeks after starting risperidone treatment. All patients were of Tamil origin, excluding the potential confounding effects of mixed ancestry. To further minimize the influence of confounding factors, our analysis was corrected for sex, age and tobacco use. We aimed to identify specific methylation sites that may help predict the AP response, but also to shed light on the functions and processes associated with variation in treatment efficacy and thus contribute to a better understanding of the broad epigenetic mechanisms underlying the risperidone treatment response.

**Methods**

The methods are described in detail in the supplementary material.

*Recruitment, sampling design and treatment response assessment*

The cohort consists of 28 Tamil speaking adults, diagnosed with schizophrenia (total score at Positive and Negative Syndrome Scale, PANSS ≥30) and newly prescribed risperidone (4-8 mg/day). Patients who were prescribed antipsychotics other than risperidone, enganged in substance abuse and pregnant or lactating women were excluded from the study. Treatment compliance was confirmed by the patients' family members or caregivers. None of the patients received any antipsychotic medication at least four weeks before the baseline visit. All except one were comedicated with at least one additional medicament for sleep or to alleviate side effects. Good treatment response was defined as at least 20% reduction of total PANSS score after four weeks of treatment [32]. Blood samples for the methylome analysis were taken during baseline and follow-up visits.

Patients' characteristics summary, including the differences between the good and bad responders can be found in Table 1.

*Ethical statement*

Informed consent was obtained from each study participant and the patient's legally acceptable representative (LAR) or a family member. The Institutional Ethics Committee approved the study protocol before the commencement of the study (IEC Project No. JIP/IEC/4/2013/189).

*Generation and quality control of methylation data*

DNA was extracted from the blood cells using QIAamp Blood Midi Kit (Qiagen, Germany) and subsequently used for bisulfite conversion with EpiTect Fast 96 Bisulfite Kit (Qiagen, Germany). Genome-wide methylation was assessed using Infinium Methylation EPIC BeadChip [33] on the Centre National de Recherche en Génomique Humaine (CNRGH) automated platform (Illumina Inc., USA). All steps were performed following the manufacturers' protocols.

Quality control, preprocessing, statistical analysis and data visualization were all done in R statistical framework [34], version 4.1.3. We first filtered out low quality probes, as they can negatively influence the normalization procedure [35], and checked sample quality [36]. In addition to low quality, probes, we also removed previously identified 44,570 cross-reactive probes (binding to multiple sites in the genome [45,46] ). Finally, to reduce the amount of technical variation within a single array, we performed a three-step (background, dye-bias and design-bias correction) normalization of the data. [50].

*Estimation of blood cell type composition and batch effects*

We estimated the relative abundances of six leukocyte types from the methylation data with the *flow.sorted.blood.EPIC* v. 1.10.1 package [40]. as the leukocyte composition is an important source of DNAm variability (for review see [41]). As the blood cell types *per se* were not of primary interest in this study and six additional variables could negatively affect inference, we used the first principal component (PC1, explaining 84.6% of variability, Supplementary Figure 1) of the PCA based on the leukocyte composition data to represent the cell type heterogeneity.

Similarly, to avoid overfitting due to a high number of independent variables and to prevent false positives due to a tendency of multilevel factors to appear significant in the statistical analysis, we we used the two "unwanted variation" ordination axes as a proxy for batch effects in the models. The batch effects included a "control matrix", as described in Fortin et al. [42], chip identity, position on the chip and sampling centre (Supplementary Figure 2).

*Multivariate analysis*

In order to examine global blood methylation patterns, we performed a PCA on all 56 samples, using variable (SD of β-values distribution < 0.1) autosomal probes only (N = 39,239), to reduce computational burden and the effect of noise [35]. This initial PCA (Supplementary Figure 3) revealed two outlier individuals (P9 and P14) that were excluded from further analysis. Of note, these two individuals had nothing conspicuous in common (they differed by sex and response to treatment from one another, had distinct DNAm profiles according to the PCA and did not differ from other patients either by duration of psychosis or comedication). In both cases, the samples from the same individual grouped together, indicating that the lack of similarity to other samples was probably due to unknown sources of biological variation (e.g. genetics or physiological condition). All the reported results are based on the remaining 52 samples from 26 individuals (19 good and 7 bad responders).

To assess the difference in magnitude of medication-induced methylation changes between the good and bad responders, we compared the pairwise Mahalanobis (multivariate) distances between the paired samples by linear regression. In order to extract CpGs significantly correlated (absolute correlation values ≥0.2) with the main axes (PC1 and PC2) of methylome variation and to assess the relationship between the risperidone response and other factors (baseline and follow-up PANSS scores, visit, sex, smoking, alcohol use, blood cell type composition, chip number, position on the chip, patient identity, unwanted variation) and those axes, we calculated the corresponding correlations using *dimdesc* function from *FactoMineR* [43].

*Gene set enrichment analysis*

In order to gain insight into the function of genes associated with the CpGs significantly correlated with either positive or negative sides of the two main principal components (PCs), we conducted an enrichment analysis for each of the four groups of CpGs separately (i.e. PC1-, PC1+, PC2-, PC2+, Figure 1D) using the *gometh* method, adapted to the biases specific for the methylation data [61] . We used only the variable 39,239 probes as a CpG universe to avoid sample source bias [45].We tested for the enrichment of Kyoto Encyclopedia of Genes and Genomes (KEGG) pathways and gene ontology (GO) categories. Both databases contain a large number of categories/pathways to be tested, strongly reducing the likelihood for any of the p-values to remain significant at any of the commonly used significance cutoffs after the multiple testing correction in the small datasets. However, gene-set enrichment analysis is a multi-step procedure and some authors argue that a p-value adjustment does not correctly account for the family-wise error rate and advise to use raw p-values and term ranking to select potentially interesting pathways and processes instead [46]. Therefore, we kept the terms/pathways with raw $p \leq 0.01$ and additionally filtered the results based on the number of differentially methylated genes and pathway/term size and coverage. In order to simplify the interpretation of the GO enrichment analysis results, we used the *GOxploreR* v. 1.2.4 package to prioritize GO terms [47] and *GOSemSim* v. 2.18.1 package [48,49] to group them based on their semantic similarity (0.7 similarity threshold). In addition, we performed a gene-set enrichment analysis of the variable 39,239 autosomal CpGs with all probes on the chip that passed our quality control step (N = 821,287) as the universe in order to identify functional categories that were over-represented in the variable portion of the blood methylome of our cohort.

*Association between individual CpG methylation levels and risperidone response*

To analyze the relationship between the risperidone response and methylation at the probe level, , we ran generalized linear mixed models with beta family (*glmmTMB* v. 1.1.2.3 package [50]) using visit, response (defined either as binary variable or %PANSS improvement) and visit x response interaction as variables of interest. We included leukocyte composition, sex, smoking and two variables describing the unwanted variation as covariates. We removed the models with convergence, multicollinearity, outlier, uniformity and dispersion issues (*DHARMa* v.0.4.6, [51]). We compared these models to the corresponding null models with a likelihood ratio test (LRT) to obtain a model significance and subsequently adjusted the p-values for multiple testing using the Benjamini-Hochberg procedure. However, due to the exploratory nature of our analysis, we included all the models with raw p-values ≤ 0.05 in the results. Still, we clearly indicate which models remained significant at 0.05 level following the Benjamini-Hochberg (B-H) correction. In addition, we added the information about the experimental evidence of correlation between the blood and brain methylation values of these CpGs from the IMAGE-CpG tool [52]. Finally, we used the top significant CpGs (absolute value of coefficient estimates from glmmTMB ≥0.18 for visit x response interaction and top twenty CpGs for the main effect of response) to try to predict the risperidone response using the stability variable selection method (package *mboost* v. 2.9-5 [53,54]).

*Data and code availability*

All data including R objects and figures that are not included as a supplementary material are available on Figshare. Raw methylation intensity data are deposited in the ArrayExpress repository under the accession number E-MTAB-11921. All scripts necessary to reproduce the presented analysis are available upon request from the authors.

**Results**

*Global methylation patterns differ between good and bad responders and are more influenced by medication in good responders*

PCA of variable autosomal probes revealed small, but noticeable differences in blood methylation patterns between good (N = 19) and bad (N = 7) responders at baseline and after four-week risperidone therapy (Figure 1A and 1B). The methylomes from the same individual were, to a large extent, temporally stable, but we did observe slight changes following the treatment (Figure 1C). Although not statistically significant, these shifts were more pronounced in good responders, as reflected by higher Mahalanobis distances between the corresponding paired samples (mean ± SD = 0.73 ± 0.524) compared with those from the bad responders (mean ± SD = 0.36 ± 0.244; Anova: $F_{1,24}$ = 3.16, p = 0.088, Figure 1C).

*Global methylation patterns are correlated with blood cell type composition, PANSS score and %PANSS improvement*

To identify factors that best describe variation along the first two ordination axes, we calculated correlations between these factors and the axes, filtering out those with correlation values <0.2 (Figure 1D). In addition to the variables describing technical and patient identity related variation, we found a strong negative correlation of the first principal component (PC1) with the blood cell composition PC (cor = -0.95, p<$10^{-6}$) and a moderate positive correlation with the baseline PANSS score (cor = 0.37, p = 0.008, Supplementary Table 1). On the other hand, the follow-up PANSS score was negatively correlated with the PC2 (cor = -0.33, p = 0.018, Supplementary Table 1) and thus the %PANSS improvement was correlated with it positively (cor = 0.29, p = 0.04). The PC1 partially separated good and bad responders (cor = 0.15, p = 0.05, Supplementary Table 1) and strongly correlated with the blood cell composition PC, probably reflecting a higher proportion of B-cells in good responders both at baseline and follow-up visits (and a marginally significant lower proportion of the CD4+ T-cells at the baseline, Table 1).

In a similar manner, we calculated correlations between the PCs and CpG methylation values, keeping again only those with the absolute correlation values ≥0.2 (Supplementary Table 2). Overall, 20,747 (53%) of the variable autosomal variables were significantly correlated with the first two PCs. Specifically, 9,986 and 2,975 CpGs were positively and negatively correlated with the PC1 (representing the groups PC1+ and PC1-, Figure 1D), and 6,326 and 3,660 CpGs were positively and negatively correlated with the PC2 (representing the groups PC2+ and PC2-, Figure 1D), respectively. Some of the genes (e.g. *RELN, GRIN2A, GRIN2B, DLG2*) associated with the CpGs significantly correlated with the PC2, and thus to a certain extent co-varying with the treatment response (% PANSS improvement or follow-up PANSS) have been previously linked to the brain development, SCZ and other psychiatric disorders (Supplementary Table 2).

### *GO enrichment analysis revealed many variable CpGs associated with the genes involved in immunity and neuron functioning and development*

GO enrichment analysis of the four tested CpG groups (PC1-, PC1+, PC2-, PC2+, Figure 1D) revealed the enrichment of biological processes (BP) related to immunity, neurogenesis and synaptic signaling, in addition to more general regulatory and metabolic processes (Supplementary Table 3). Specifically, the CpGs with a tendency towards higher methylation values in bad responders and a lower relative abundance of B-cells (i.e. PC1-) were annotated to the genes primarily involved in different aspects of immunity and blood cells differentiation (e.g. positive regulation of inflammatory response, adaptive immune response, establishment of T-cell polarity), and to a lesser extent in synaptic signaling (e.g. excitatory postsynaptic potential, long-term postsynaptic potentiation), presynapse assembly, but also in ERBB signaling and amyloid fibril formation. The processes enriched along the opposite side of the PC1 (i.e. PC1+, the CpGs more methylated in the good responders) included, among other, Fc receptor signaling pathway, endosomal and intracellular transport, as well as terms related to phagocytosis, inflammatory response (interleukin pathway and neutrophil degranulation), JUN kinase activity and neuron projection development (Supplementary Table 3, Figure 2). Whereas the processes enriched along the PC1 were mainly related to immunity, those enriched along the PC2 were dominated by neuron functioning and development. Specifically, the CpGs with higher methylation values in the patients with a higher follow-up PANSS score (PC2-), were associated with the genes involved in synapse organization and functioning (e.g. postsynaptic membrane organization, neurotransmitter-gated ion channel clustering, regulation of synapse vesicle clustering), axon development (e.g. axonal fasciculation) and signaling pathways involved in development, carcinogenesis and neurogenesis (positive regulation of Notch signaling pathway, positive regulation of Wnt pathway). In addition, this group was characterized by the enrichment of processes related to gastric motility, circadian clock entrainment and cardiac muscle development (Figure 2). On the other hand, the processes associated with symptom improvement (PC2+) encompassed cognition and synaptic signaling (e.g. ionotropic glutamate receptor signaling pathway, regulation of N-methyl-D-aspartate (NMDA) receptor activity, excitatory chemical synaptic transmission) and some regulatory (e.g. positive regulation of renal sodium excretion), metabolic and developmental processes (e.g. blood vessel maturation).

We further wanted to know if the enriched biological processes were reflected in other gene ontology categories, namely in the cellular components (CC) and molecular functions (MF, Supplementary Table 3). This held largely true for CC, with the enriched terms dominated by neuron-related structures for all except the PC1+ group (e.g. Golgi apparatus subcompartment, AP-3 adaptor complex, Supplementary Figure 4), with the corresponding enriched molecular functions (e.g. MAPKKK activity, ubiquitin protein ligase activity, Supplementary Figure 5). Similarly, the molecular functions over-represented in the PC2+ group were related almost exclusively to neuronal signaling (e.g. NMDS glutamate receptor activity), whereas the the PC1- and PC2- groups were associated with more general, higher level terms (e.g. steroid dehydrogenase activity and acting binding, respectively).

Finally, we wanted to check which processes and cellular components were over-represented in the variable CpG subset that we used as input for the multivariate analysis compared to the universe containing all the probes present on the EPIC chip. Among other, these processes included: transmission of nerve impulse, transport across blood-brain barrier, neuron projection morphogenesis, regulation of neurotransmitter receptor activity, dopamine receptor signaling pathway, modulation of chemical synaptic transmission myeloid leukocyte mediated immunity and leukocyte degranulation (Supplementary Figure 6). The enrichment of neuron-functioning related processes was further reflected in the enriched cellular components, to a large extent associated with neurons (neuronal dense core vesicle, axon, presynapse, dendrite, neuronal cell body etc.). This suggests that a significant part of epigenetic variation observed in this cohort is related to immunity and nervous system functioning.

***Good response to medication is associated with hypermethylation of CpGs related to nicotine and morphine addiction KEGG pathways***

In a corresponding KEGG pathway analysis, we identified several immunity, cancer and infection related pathways (Supplementary Table 4) associated with either pole of the PC1. More particularly, the PC1- group was characterized by the pathways involved in adaptive immune response (e.g. T cell receptor signaling), endometrial and small cell lung cancer and infection (toxoplasmosis and Epstein-Barr virus infection), whereas the PC1+ group was characterized by *Yersinia* infection and lysosome pathways. In addition, chemokine signaling pathway, involved in inflammatory responses and general functioning of the immune system, was enriched in both PC1-related groups. Some differentially methylated genes from this pathway were detected in both groups, suggesting that the phenotypic variability associated with the PC1 could be partially explained by gene-specific epigenetic mechanisms only. However, most of the genes were detected in either PC1- or PC1+ group, indicating the existence of additional mechanisms acting on different components of these pathways. Regarding the PC2, no pathways were over-represented in the PC2- group, whereas the PC2+ group was characterized by nicotine and morphine addiction pathways (Supplementary Figure 7).

***Differentially methylated CpGs related to risperidone response are concentrated in gene bodies and outside of CpG islands***

Although we examined the relationship between individual CpG methylation values and response to medication for both %PANSS improvement and binary (good/bad) response, we focused on the latter in the rest of the analysis for simplicity. Our choice was supported by a strong correlation of the coefficient estimates for both definitions of response in the corresponding models (Spearman's rho between the %PANSS improvement and binary response for the response main effect = 0.78 [0.76, 0.79] and response x visit interaction = 0.76 [0.74, 0.78]).

Probewise analysis of CpGs from the genes associated with the enriched GO BP terms or KEGG pathways revealed 216 CpGs assigned to 193 unique genes (or 194 CpGs in 178 genes after BH correction) that were on average significantly differentially methylated between the good and bad responders regardless of the visit (i.e. response main effect, Supplementary Table 5). The majority of these probes were situated in the gene bodies (64%) and outside (68%) of the CpG islands and shores (i.e. in the "open sea" areas). Those proportions were different from those observed for all the probes on the EPIC chip, with more probes than expected situated outside the CpG islands and in 5'-UTR regions (Fisher's exact test simulated p-value = 0.0005 and 0.03, respectively, Figure 3A and Figure 4A). As a gene can be represented by multiple CpGs on the EPIC chip, we also wanted to determine the proportion of differentially methylated CpGs related to each of the above 193 genes. We found that a large majority of CpGs (97.5% on average, min = 72%, max = 99.9%) annotated to these genes were not differentially methylated between the good and bad responders. Some of the CpGs with the highest estimated methylation difference between the good and bad responders (Supplementary Table 5 and Supplementary Figure 8, Table 2) were characterized by a higher methylation in good responders (e.g. those associated with the genes *COL13A1*, *SORCS2*, *CASZ1*, *HDAC4* and *HTT*), whereas others had higher methylation values in bad responders (e.g. those annotated to the *RB1, AKAP6, PKHD1. DISP3, ABLIM2, SPRED2, BMP7* or *GSTM1* genes). Some of these genes are involved in neuron differentiation (*DISP3, ABLIM2*), while others have been linked to neuromuscular disorders (*COL13A1*), Huntington disease (*HTT, SORCS2*), schizophrenia (*AKAP6*) and drug detoxification (*GSTM1*). However, none of these or any other of the top 20 differentially methylated CpGs were kept as a reliable predictor of the risperidone response by the stability variable selection procedure (maximum selection probability 0.46 for a *COL13A1*-related CpG for baseline methylation values model and 0.52 for *CASZ1* CpG for both visits model, Table 2, Supplementary Table 6).

In a corresponding analysis, we identified 237 CpGs in 221 genes (or 208 CpGs in 194 genes after B-H correction) with distinct temporal dynamics in good and bad responders (i.e. with a significant response x visit interaction, Supplementary Table 5 and Supplementary Figure 9).

Again, the majority of these probes were situated more often than expected by chance in the open sea areas (64%, Fisher's exact test p-value = 0.0005, Figure 3B-C) and concentrated in the gene bodies (69%, Fisher's exact test p-value = 0.001, Figure 4B-C) . Once more, these differentially methylated CpGs represented only a minority of the total CpGs on the chip associated with their corresponding genes (2.5% on average, min = 0.13%, max = 33%). Only four of these CpGs, annotated to the *NAV2, MGST1, BPGM* and *CACNA2D2* genes, were represented by good quality models with the absolute values of *glmmTMB* coefficient estimates ≥0.18. Each of those CpGs exhibited a specific pattern of change. Methylation of the *NAV2*-associated CpG, a gene involved in cellular growth and migration, and of *CACNA2D2*-related CpG, involved in opioid and gabapentin metabolism, decreased in bad responders and increased slightly in good responders following the treatment. On the other hand, the methylation of the CpG associated with *MGST1*, a detoxification enzyme, increased in bad responders and decreased in the good ones, converging at a similar level at the follow-up visit for both groups. The remaining CpG was associated with *BPGM* (a gene influencing erythrocyte oxygen affinity) was more strongly methylated in good responders at the baseline, with the values dropping to the stable bad-responder levels at the follow-up. Stability selection of the variables potentially predictive for the treatment response identified the *CACNA2D2*-associated CpG as a good candidate (0.82 selection probability), followed by *NAV2* (0.52. Table 3, Supplementary Table 6). However, note that the model included both baseline and follow-up methylation values for those CpGs, and it was not possible to predict the response from the baseline values only.

*Discussion*

SCZ patients vary widely in their response to AP treatment and finding a suitable medication in individual cases may take a long time. In this study, we explore blood methylomes of 28 SCZ patients before and after risperidone treatment in search for epigenetic signatures that might help predict the treatment outcome. Although we could not identify any particular CpG whose methylation values would predict response to risperidone treatment, we detected small but noticeable differences in global methylation patterns between good and bad responders and we observed little change between the baseline and follow-up visits for either group. The two groups were in addition characterized by temporally stable, but different blood cell composition, notably by a higher proportion of B lymphocytes in good responders. Finally, the functional analysis of significantly differentially methylated probes in our cohort suggests that the level of DNA methylation of genes involved in immunity, neurotransmission, and neuronal development could play a role in AP response.

Previous research has shown that SCZ patients sometimes have higher neutrophil-to-lymphocyte ratio (NLR) compared to healthy controls and revealed its positive correlation with symptom severity as well as NLR decrease in medicated patients (review in [55]). If and how the percentage of B-cells is related to SCZ pathology and treatment response remains unclear, although some studies reported a decrease of the B-cells percentage upon AP treatment (review in [56]). We detected no change of neutrophil relative abundance at the follow-up, nor any correlation with PANSS score (not shown), but we did find marginally lower neutrophil proportion in good responders as well as consistently (on both visits) lower percentage of B-cells in bad responders, suggesting that immune status and inflammation could be important for predicting AP response. However, note that we only have indirect blood cell composition estimates based on methylation data and not absolute counts, which are necessary for accurate description of leukocyte populations. The role of inflammation, stress and infection in SCZ has been recognized for a long time (review in [78,79].). Similarly, some evidence suggests that the levels of certain cytokines may help predict the outcome of SCZ treatment [59,60]. In our cohort, we observed response-specific changes following the treatment for three interleukin receptor-associated genes – *IL23R, IL1R1* and *IL18RAP1* - as well as *IFNL1* cytokine (Supplementary Table 5), and enrichment of chemokine signaling, Th1 and Th2 cell differentiation and several cancer and infection (including toxoplasmosis) pathways related genes (Supplementary Figure 7, Supplementary Table 4). Increased titers of *Toxoplasma gondii* antibodies have been linked to both acute and chronic psychosis [61] and our results suggest that infections might affect the response to AP just as they affect the SCZ etiology, by driving DNA methylation changes. However, it is so far unclear how the methylation shifts observed here translate into gene expression and physiological processes, as the relationship between methylation and gene expression depends on various factors, such as tissue, CpG position and expression level of the gene in question [62,63].

It has been shown that genetic variation affects epigenetic mechanisms in SCZ [64,65]. Although we do not have genetic data to test the relationship between genetic polymorphisms, methylation patterns and AP response explicitly, we detected differences in methylation of CpGs related to the genes previously associated with SCZ treatment. For example, symptom improvement in our study was associated with glutamate receptor signaling pathway and NMDA receptor activity, just as differential methylation of the related genes and SNPs in glutamate receptors have been linked to AP treatment outcome by both candidate gene [14] and genome-wide [66] approach previously. Similarly, we detected differential methylation of MHC-related genes (i.e. *HLA-DQ1*, *HLA-DRB1* and a pseudogene *HLA-H* in our study (Supplementary Table 5), which also affect response to AP drugs [67]. While the exact genes detected in our study do not correspond to the genes reported previously, the functional overlap suggests that genetic and epigenetic mechanisms involving the same biological processes contribute to the AP treatment outcome. The lack of perfect match in this case could be explained by the choice of method (candidate vs. whole genome approach), as well as other biological (medication, patient characteristics) and technical (e.g. study size) factors.

Although polymorphisms within cytochrome P450 gene *CYP2D6* have previously been linked with variation in risperidone response (for review see [68]), we found no evidence of differential methylation of any of the nine CpGs associated with this gene on the EPIC Infinium chip. On the other hand, three CpGs in the glutathione S-transferase *GSTM1* (Table 2) were among the most differentially methylated between the good and bad responders, whereas a CpG associated with another glutathione S-transferase *MGST1* (Table 3) was among the probes exhibiting the most pronounced response-specific temporal dynamics. Glutathione S-transferases are involved in xenobiotic metabolism and could therefore also directly shape antipsychotic response. Notably,

Pinheiro et al. [69] found that polymorphisms in *GSTM1* affect the risk of treatment-resistant schizophrenia, whereas polymorphisms in *MGST1* have been associated with the side effects (sinus bradycardia) of second-generation antipsychotics in Chinese population [70]. Glutathione S-transferases could therefore also present a potentially interesting target for future studies of AP-response predictive biomarkers.

So far, few studies examined epigenetic mechanisms underlying AP response and even fewer focused on methylation profiles of human peripheral tissues, which have potential as a source of clinically relevant biomarkers (reviews in [22,23]). For example, methylation of serotonergic pathway genes has been implicated in SCZ etiology, as well as in response to AP [71] using targeted approach, but we found no evidence for this in our data. Similarly, we did not detect differential methylation of previously examined dopaminergic system genes (*ANKK1* and *DRD2* [72]), nor any other genes identified by genome-wide approach [25]. On the other hand, we found several differentially methylated procadherin and cadherin genes, corroborating the results of a study of olanzapine-induced changes in rat brains [73]. These discrepancies are likely due to a number of different factors such as small sample size (making it hard to detect presumably small-effect differences contributing to the treatment outcome [15]), known and unknown biological variation including ethnicity [74], heterogeneity of experimental conditions or different APs and their combinations used in these studies [75]. In addition, most studies focused on specific genes or groups of genes [71–73], with the whole methylome approach remaining an exception ([25], this study). Finally, apart from our study, only Rukova et al. [25] applied a repeated-measures experimental design, whereas others focused either on differences before or after treatment.

Overall, we detected a number of genes whose methylation is potentially relevant to AP response, many of which are involved in neurotransmission, neuronal development and functioning and other processes that have been previously implicated in SCZ etiology. For example, *HDAC4* methylation differed between the good and bad responders both on average and by its temporal dynamics, indicating epigenetic mechanisms involved in histone acetylation may influence the treatment outcome as well. On the other hand, we were not able to predict the outcome from the baseline methylation values of any of the genes identified as significantly affecting response in the probewise analysis (Table 2, Supplementary Table 6). However, if we included baseline and follow-up methylation values – in other words, methylation temporal dynamics – we identified one CpG, annotated to *CACNA2D2*, encoding the Calcium Voltage-Gated Channel Auxiliary Subunit Alpha2delta 2, as with a possible predictor of AP response (Table 3, Supplementary Table 6). Calcium voltage-gated channels play role in neurotransmission and neuron plasticity and polymorphisms in their genes have been associated with SCZ risk and pathogenesis [76–78]. We identified an additional gene, NAV (neuron navigator 2), whose methylation temporal dynamics was moderately correlated with AP response. [96]. Polymorphisms in *NAV2* gene, essential for brain development, affect onset of Alzheimer disease [80] and associated episodic memory phenotypes [81]. To summarize, functions and associations of these genes and other candidates, for example *RNF19A* (involved in reduced adult neurogenesis [82]), *SHANK2* (linked with mania-like behavior [83] and autism [84]), *LMF1* (implicated in metabolic disorders [85]) or those significantly indirectly correlated with the symptom improvement only in the multivariate, but not in the probewise analysis (e.g. *RELN*, *DLG2*, *CUX2*), cover various aspects of SCZ etiology and manifestations and support the hypothesis that response to AP treatment is complex and shaped by multiple interacting processes and underlying epigenetic mechanisms.

At this point, we would like to address the limitations of our study and some general issues that require further attention in the future. The small size of our cohort (twenty good and eight bad responders, with one biological outlier in each of these groups that were therefore removed from the analysis, Supplementary Figure 3) does not allow for strong conclusions, especially as our results suggest that there is no single (or few) large-effect CpG site predicting the outcome of SCZ treatment. We can assume that the size of the cohort at least partially explains why we were not able to predict the response from the baseline methylation values. Other factors that might have affected our ability to detect conclusive correlations between methylation and AP response include variable duration of psychosis, inflammation status [57], as well as comedication [86] of the patients in our cohort. Furthermore, correlation between blood and brain methylation values for most of the genes in our studie is small or unknown [52]) and even so, the effect of methylation on gene expression level is not straightforward [21,62,87]. Despite these issues, we were able to demonstrate small global differences between the blood methylomes of SCZ patients varying in their response to medication and found that many of the implicated genes and processes have been previously associated with genetic basis of both SCZ pathology and AP response. In addition, monitoring of temporal methylation dynamics of *CACNA2D2*-associated CpGs after starting medication could enable more rapid switch of therapy where needed. However, in order to improve our search for predictive biomarkers and address the limitations listed above, we call for longitudinal study designs with bigger and well characterized cohorts (inflammatory markers, smoking, sex, age, ethnicity...), including validation sets, preferably with the same treatment and psychosis duration [24]. In addition to longitudinal probing of SCZ patients, the inclusion of healthy controls could contribute to understanding and interpretation of methylation changes observed after treatment and linking them to clinical outcomes. Finally, we need to improve our knowledge about the relations between the epigenomes from different tissues as well as about the relationship between genetic and epigenetic variation in SCZ etiology and AP pharmacodynamics [23].

**Conclusion**

We present here the first (to the best of our knowledge) repeated-measures study aiming at better understanding and prediction of the antipsychotic (AP) treatment response from the blood methylomes of schizophrenia patients with a genome-wide coverage (including both CpG islands and open sea areas). Overall, response heterogeneity seems to be shaped by both stable differences in methylation levels as well as by response-specific shifts following the treatment start. Despite the small sample size, the genetic uniformity and detailed description of our cohort allowed us to identify potential processes (mainly related to immunity and neuronal functioning) involved in the AP response and suggest potential targets, such as *CACNA2D2* or glutathione S-transferase *MGST1* and *GSTM1* genes, for the future biomarker development. We look forward to the future studies with larger cohorts and thus the power to identify additional potential epigenomic biomarkers of antipsychotic response as well as to examine the utility of those identified in our work.

**Declaration of interest**

The authors declare they have no competing financial interests or personal relationships that could influence the reported work.

**Acknowledgements**

The study was supported by the Investissements d'Avenir program managed by the Agence Nationale pour la Recherche (ANR) under reference ANR-11-IDEX-0004-02 (Labex BioPsy), ANR-10-COHO-10-01 (Cohorte PSY-COH) and Indo-French Centre for the Promotion of Advanced Research (IFCPAR/CEFIPRA) for promoting the collaboration (Grant No. IFC/4098/RCF2016/853) between Jawaharlal Institute of Postgraduate Medical Education and Research (JIPMER) and University of Sorbonne universities, Paris, France. This work also received financial support from the Institut National pour la Santé et la Recherche Médicale (Inserm), the Fondation FondaMental, the International Foundation (IF), and the Commissariat à l'énergie atomique et aux énergies alternatives (CEA). The Jamain's team is affiliated with the Paris School of Neuroscience (ENP) and the Bio-Psy Laboratory of Excellence.

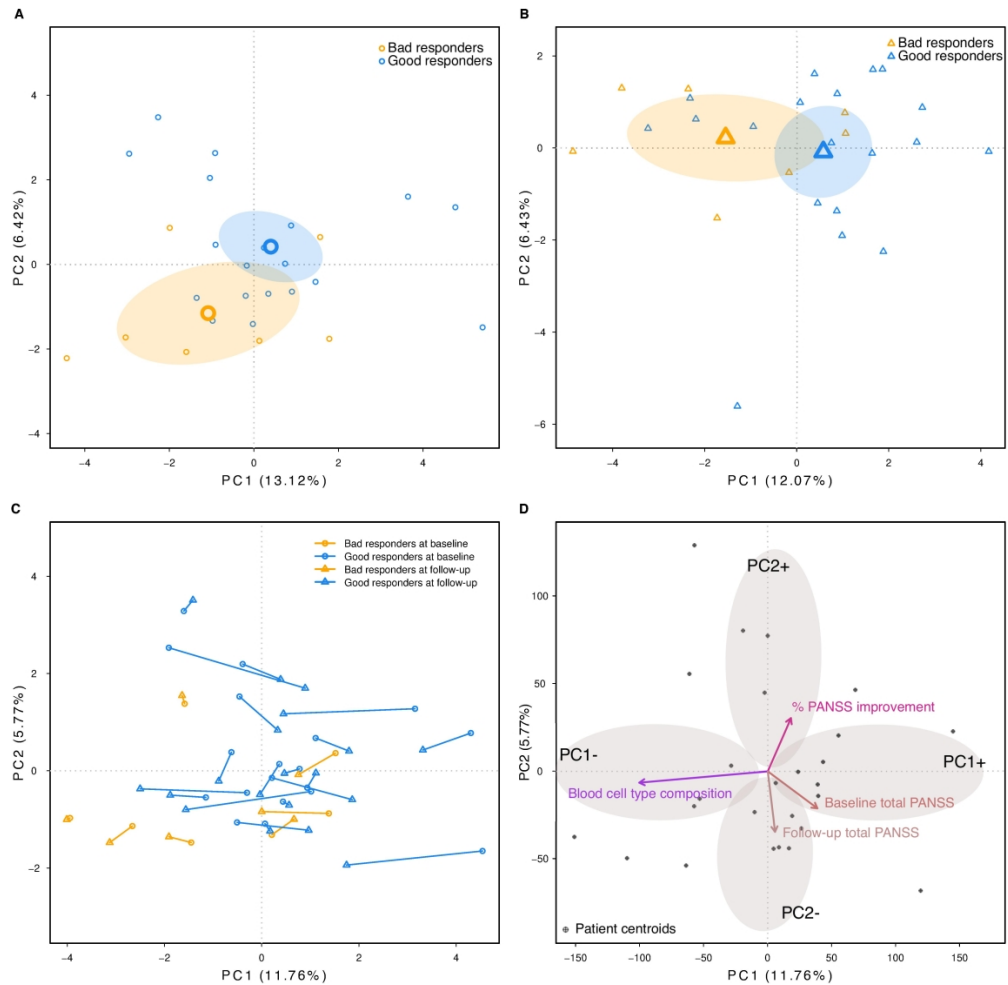

Figure 1. PCA of within-normalized methylation values of variable autosomal probes without outlier individuals. The panels show individual samples at the A) baseline, B) follow-up and C) both visits together. D) The same PCA as in C), but showing factors with the correlation of at least 0.2 with any of the two depicted PCs. PC-correlated CpGs, PANSS score components and unwanted variation variables are omitted for simplicity (complete results can be found in the Supplementary Table 2 & 3). On A) and B), big plotting characters denote centroids, the ellipses the 95% confidence regions for centroids locations. On C), the lines connect the baseline (dot) and follow-up (triangle) samples from the same individual. On D), the ellipses show app. areas that contain variables and CpGs significantly positively correlated with the corresponding side of the axis and therefore belonging to one of the four groups defined in the Results (PC1-, PC1+, PC2-, PC2+).

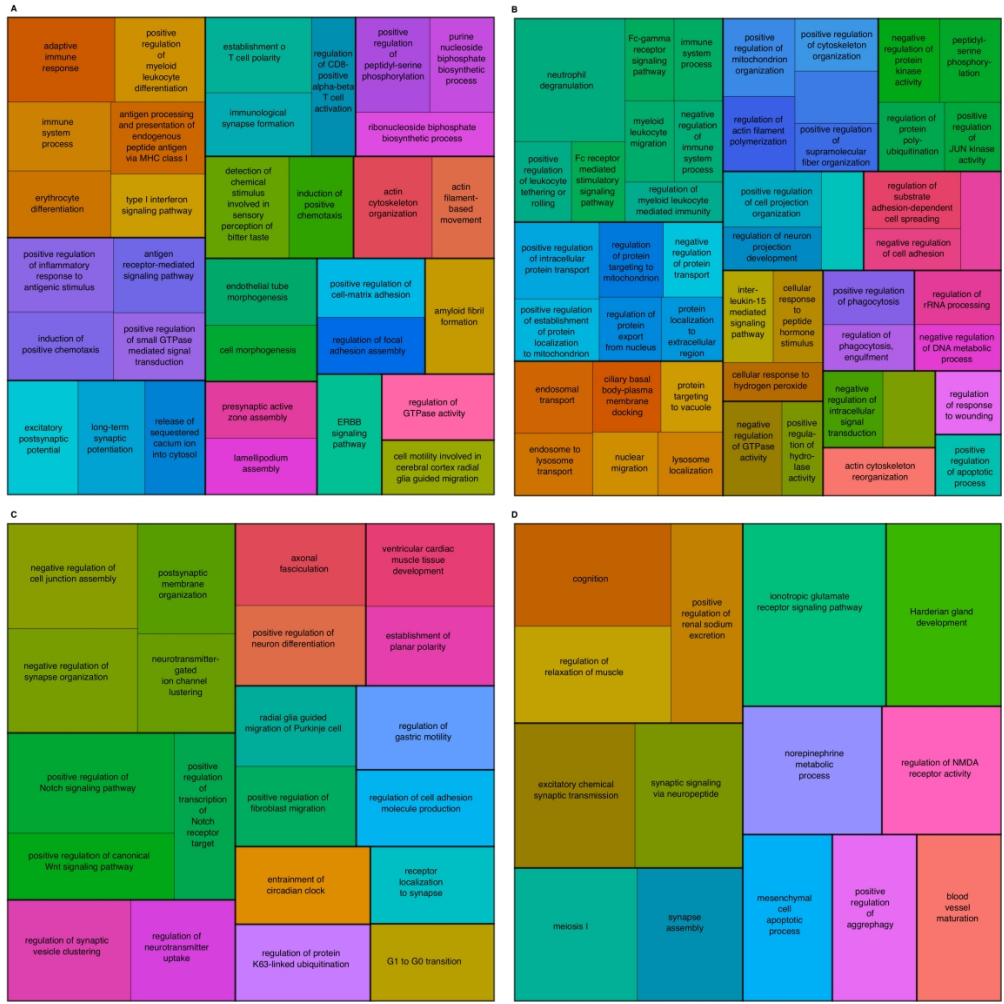

Figure 2. Treemaps of prioritized GO BP terms. The GO terms enriched in A) PC1-, B) PC1+, C) PC2- and D) PC2+ groups (s. Figure 1D) are grouped by semantic similarity at 0.7 cutoff. The size of rectangles is proportional to the significance of the corresponding term. Sets of semantically similar terms are divided from each other by thick border lines.

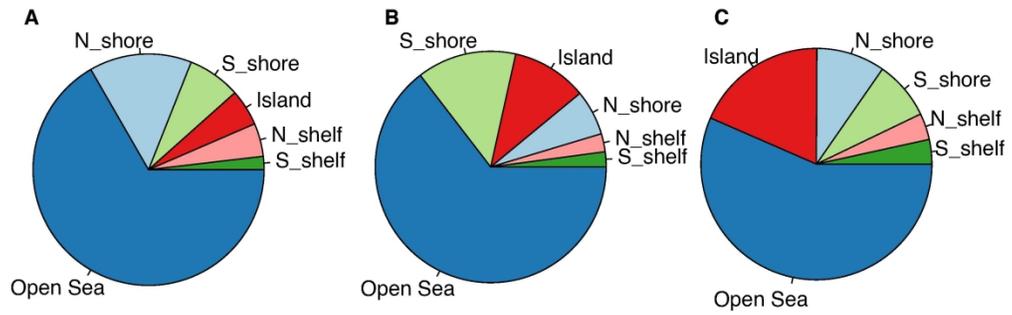

Figure 3. Pie charts showing the distribution of probes in relation to CpG islands. CpGs significant for the A) main effect of response, B) response x visit interaction and C) all CpGs on the EPIC chip.

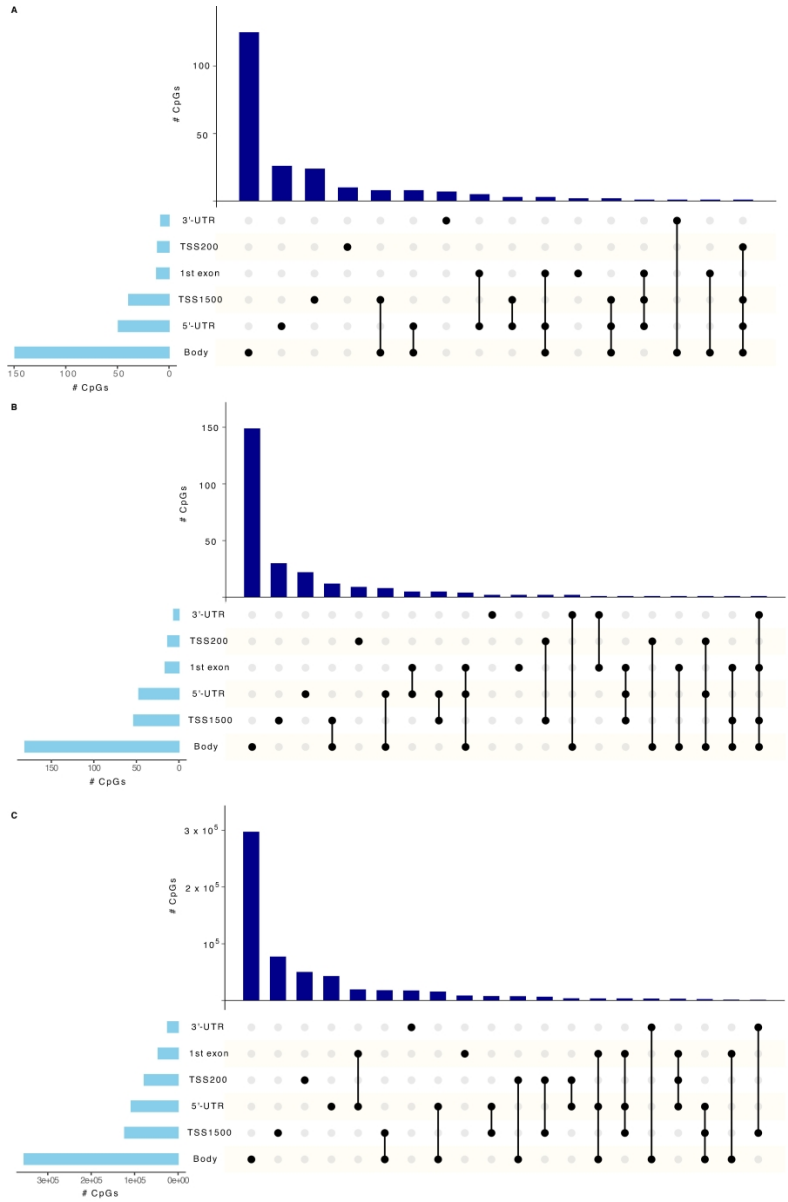

Figure 4. Upset plots showing the distribution of probe positions in the genes. CpGs significant for the A) main effect of response, B) response x visit interaction and C) all CpGs on the EPIC chip.

**Table 1**. Cohort description including the differences between good and bad responders. Means are shown with corresponding standard deviations.

| | Bad responders (N=8) | Good responders (N=20) | Test | Stat | p-value | Effect size |
|---|---|---|---|---|---|---|
| % males | 75 | 50 | Fisher's exact test | | 0.40 | |
| Age (years) | 34.7 ± 12.31 | 32.2 ± 7.16 | Welch Two Sample t-test | t = 0.55, df = 8.96 | 0.60 | |
| Weight (kg) | 58.9 ± 15.91 | 55.7 ± 14.89 | Welch Two Sample t-test | t = 0.49, df = 12.22 | 0.64 | |
| Past or current alcohol user (%) | 12.5 | 20 | Fisher's exact test | | 1 | |
| Past or current smoker (%) | 25 | 30 | Fisher's exact test | | 0.28 | |
| Duration of psychosis (months) | 19.1 ± 22.51 | 35.9 ± 30.25 | Welch Two Sample t-test | t = -1.61, df = 17.42 | 0.12 | |
| Total PANSS score at baseline | 89.5 ± 17.48 | 93.05 ± 12.62 | Welch Two Sample t-test | t = -0.52, df = 10.06 | 0.61 | |
|     Positive score at baseline | 26.3 ± 5.39 | 28.3 ± 5.02 | Welch Two Sample t-test | t = -0.92, df = 12.15 | 0.37 | |
|     Negative score at baseline | 22.5 ± 6.05 | 21.7 ± 4.91 | Welch Two Sample t-test | t = 0.33, df = 10.9 | 0.75 | |
|     General Psychopathology score at baseline | 40.8 ± 7.80 | 43.1 ± 6.18 | Welch Two Sample t-test | t = -0.75, df = 10.71 | 0.47 | |
| **4-week total PANSS score improvement (%)** | **14.8 ± 4.77** | **43.0 ± 10.78** | **Welch Two Sample t-test** | **t = -9.58, df = 25.55** | **6E-10** | **0.88** |
| Risperidone dose (mg) | 5.1 ± 1.46 | 4.7 ± 1.00 | Welch Two Sample t-test | t = 0.69, df = 9.85 | 0.51 | |
| Blood cell composition (proportion at the baseline) | | | | | | |
|     Neutrophils | 0.66 ± 0.068 | 0.60 ± 0.082 | Welch Two Sample t-test | t = -1.81, df = 15.72 | 0.09 | |
|     **B-cells** | **0.05 ± 0.010** | **0.07 ± 0.020** | **Welch Two Sample t-test** | **t = -4.31, df = 25.54** | **0.0002** | **0.65** |
|     CD8+ T-cells | 0.13 ± 0.043 | 0.15 ± 0.034 | Welch Two Sample t-test | t = -1.32, df = 10.55 | 0.21 | |
|     **CD4+ T-cells** | **0.06 ± 0.014** | **0.08 ± 0.033** | **Welch Two Sample t-test** | **t = -2.13, df = 25.76** | **0.04** | **0.39** |
|     Natural killer cells | 0.07 ± 0.023 | 0.08 ± 0.020 | Welch Two Sample t-test | t = -0.84, df = 11.69 | 0.42 | |
|     Monocytes | 0.07 ± 0.020 | 0.08 ± 0.020 | Welch Two Sample t-test | t = -0.59, df = 12.97 | 0.56 | |
| Blood cell composition (proportion at the follow-up) | | | | | | |
|     Neutrophils | 0.67 ± 0.095 | 0.62 ± 0.067 | Welch Two Sample t-test | t = 1.39, df = 9.93 | 0.196 | |
|     **B-cells** | **0.05 ± 0.012** | **0.06 ± 0.018** | **Welch Two Sample t-test** | **t = -2.84, df = 19.98** | **0.010** | **0.55** |
|     CD8+ T-cells | 0.12 ± 0.041 | 0.15 ± 0.028 | Welch Two Sample t-test | t = -1.51, df = 9.58 | 0.163 | |
|     CD4+ T-cells | 0.06 ± 0.029 | 0.08 ± 0.028 | Welch Two Sample t-test | t = -1.42, df = 12.68 | 0.180 | |
|     Natural killer cells | 0.07 ± 0.019 | 0.08 ± 0.018 | Welch Two Sample t-test | t = -1.52, df = 12.74 | 0.154 | |
|     Monocytes | 0.07 ± 0.023 | 0.07 ± 0.017 | Welch Two Sample t-test | t = -0.21, df = 10.24 | 0.840 | |

**Table 2**. CpGs with the highest estimated methylation difference between the good and bad responders used for risperidone response prediction. The CpGs are ordered by the absolute value of the coefficient estimate (log odds ratio). Raw p-values and adjusted p-values represent the significance of the model including response, visit, their interaction and the covariates compared with the null model by Likelihood Ratio Test (LRT). Last column shows selection probabilities for the prediction of risperidone response for all selected CpGs or the maximum selection probability in case none of the CpGs was selected.

| CpG | Gene | Relation to CpG island | Model p-value (LRT) | Model adj. P-value (LRT) | Response main effect (log odds ratio) | Response main effect 95% confidence intervals | Stability selection (cutoff = 0.7) |
|---|---|---|---|---|---|---|---|
| cg17296678 | COL13A1 | OpenSea | 0.004 | 0.010 | -1.726 | (-2.564, -0.887) | Baseline only model: 0.46 |
| cg15108060 | RB1 | OpenSea | 0.029 | 0.051 | 1.301 | (0.486, 2.115) | |
| cg20270941 | SORCS2 | OpenSea | 0.046 | 0.075 | -1.296 | (-2.519, -0.072) | |
| cg09998151 | AKAP6 | OpenSea | 0.033 | 0.056 | 1.185 | (0.157, 2.213) | |
| cg26820259 | PKHD1 | OpenSea | 0.005 | 0.012 | 1.156 | (0.339, 1.972) | |
| cg07936689 | DISP3 | N_Shore | 0.018 | 0.034 | 1.155 | (0.416, 1.895) | |
| cg16020483 | CASZ1 | S_Shelf | 0.024 | 0.044 | -1.121 | (-1.868, -0.375) | Both visits model: 0.52 |
| cg13219409 | HDAC4 | OpenSea | 0.001 | 0.002 | -1.097 | (-1.7, -0.493) | |
| cg26128129 | HTT | OpenSea | 0.000 | 0.000 | -1.070 | (-2.125, -0.014) | |
| cg14872707 | ABLIM2 | OpenSea | 0.012 | 0.026 | 1.015 | (0.254, 1.776) | |
| cg25845314 | SPRED2 | OpenSea | 0.045 | 0.074 | 1.011 | (0.207, 1.815) | |
| cg20336007 | BMP7 | OpenSea | 0.008 | 0.018 | 1.004 | (0.038, 1.97) | |
| cg18938907 | GSTM1 | Island | 0.000 | 0.000 | -0.937 | (-1.575, -0.3) | |
| cg17901463 | GSTM1 | Island | 0.003 | 0.008 | -0.909 | (-1.524, -0.294) | |
| cg16180556 | GSTM1 | Island | 0.018 | 0.035 | -0.884 | (-1.71, -0.058) | |
| cg02249911 | | S_Shore | 0.027 | 0.049 | -0.870 | (-1.275, -0.466) | |
| cg17171259 | ROR1 | OpenSea | 0.000 | 0.000 | 0.856 | (0.043, 1.668) | |
| cg19759847 | MKS1 | OpenSea | 0.001 | 0.002 | 0.816 | (0.157, 1.475) | |
| cg23876203 | HOXA3 | N_Shelf | 0.050 | 0.081 | 0.810 | (0.185, 1.436) | |
| cg15979214 | TRIM10 | OpenSea | 0.000 | 0.000 | 0.808 | (0.02, 1.595) | |

**Table 3**. CpGs with the highest estimated differences in temporal dynamics between the good and bad responders. The CpGs are ordered by the absolute value of the coefficient estimate (log odds ratio). Raw p-values and adjusted p-values represent the significance of the model including response, visit, their interaction and the covariates compared with the null model by Likelihood Ratio Test (LRT). Last column shows selection probabilities for the prediction of risperidone response for all selected CpGs or the maximum selection probability in case none of the CpGs was selected.

| CpG | Gene | Relation to CpG island | Model p-value (LRT) | Model adj. P-value (LRT) | Visit x response interaction (log odds ratio) | Visit x response interaction 95%CI | Stability selection (cutoff = 0.7) |
|---|---|---|---|---|---|---|---|
| cg23501051 | NAV2 | S_Shore | 0.001 | 0.002 | 0.237 | (0.113, 0.362) | |
| cg25377680 | MGST1 | OpenSea | 0.001 | 0.003 | -0.228 | (-0.413, -0.042) | Baseline: 0.27 |
| cg11334105 | BPGM | OpenSea | 0.001 | 0.003 | -0.208 | (-0.394, -0.023) | |
| cg09451427 | CACNA2D2 | OpenSea | 0.018 | 0.035 | 0.181 | (0.086, 0.276) | Both visits: 0.84 |